# Transparent TiO$_2$ Nanotubes Arrays Directly Grown on Quartz Glass Used in Front- and Back-Side Irradiation Configuration for Photocatalytic H$_2$ Generation

JeongEun Yoo,[a] Marco Altomare,[a] Mohamed Mokhtar,[b] Abdelmohsen Alshehri,[b] Shaeel A. Al-Thabaiti,[b] Anca Mazare,[a] and Patrik Schmuki[a,b]*

[a] Department of Materials Science, Institute for Surface Science and Corrosion WW4-LKO, Friedrich-Alexander University, Martensstraße 7, D-91058 Erlangen, Germany;
[b] Chemistry Department, Faculty of Sciences, King Abdulaziz University, 80203 Jeddah, Saudi Arabia Kingdom;
* Corresponding author. E-mail: schmuki@ww.uni-erlangen.de.

**A B S T R A C T**

In the present work we explore the front- and back-side performance of a photocatalytic platform consisting of self-organized $TiO_2$ nanotube layers formed by complete anodization of Ti metal films evaporated on quartz slides. The adhesion and light transmission of the tube layers on the quartz surfaces are optimized by a suitable anodization procedure. After their growth, the nanotube arrays were converted into crystalline structures and sputter-coated with co-catalytic Pt nanoparticles. The optically transparent quartz substrates enable the use of the Pt-decorated tube layers for photocatalytic $H_2$ generation under either front- or back-side illumination configurations. The nanotube films on quartz are characterized in view of their physico-chemical properties, including their light transmission features measured using different light sources. The results show that the front-side illumination under optimized factors, *i.e.*, the amount of loaded co-catalyst, yields a maximized photocatalytic performance in terms of $H_2$ generation.

***Keywords:*** $TiO_2$ nanotubes; Platinum; Photocatalysis; $H_2$ evolution; Light transmission

**Introduction**

Since the first report of Fujishima and Honda[1] on the photo-electrochemical properties of titanium dioxide, $TiO_2$-based materials have become one of the most investigated photo-anodes and photocatalysts in processes such as, *e.g.*, waste water purification, water splitting, and synthesis of valuable compounds.[2-6] The success of $TiO_2$ in photocatalysis is mainly due to its high chemical stability, low cost, and, most importantly, to its suitable band-edge positions that allow, in principle, for splitting water and therefore for generating $H_2$ (*i.e.*, a most important energy vector of the future). For practical use, in order to overcome kinetic limitations of $H_2$ generation on $TiO_2$, usually a co-catalyst (*e.g.*, Pt, Pd, Au) is decorated on the semiconductor surface. Additionally, to reach a high specific reaction rate (high specific surface area, short carrier diffusion path), $TiO_2$ morphologies such as $TiO_2$ nanoparticles are used.[2,7]

However, more recently one-dimensional (1D) $TiO_2$ nanostructures, such as arrays of nanowires, nanorods and nanotubes (NTs), are widely investigated as promising morphologies due to the 1D morphology that combines a relatively high surface area with a directional charge transport. Moreover, the features of 1D semiconductor architectures can be further improved by doping,[8,9] surface modification with co-catalysts or visible light sensitizers,[10,11] and by specific thermal treatments.[12,13]

A particular type of 1D $TiO_2$ structures, anodic $TiO_2$ nanotube arrays, can be easily fabricated by anodization of a piece of Ti metal, under self-organizing conditions. Their morphology (*i.e.*, length, diameter, wall thickness and degree of ordering) can be tailored by the electrochemical conditions (electrolyte composition, applied voltage, anodization time, etc.).[14,15]

Aside from the straightforward anodization of a piece of Ti, the growth of anodic $TiO_2$ NTs can successfully occur also directly on a conducting glass substrates (*i.e.*, FTO), that is, by anodizing Ti films deposited on conductive glass.[16,17] Alternatively, the tube layers formed on a Ti metal piece can be detached as self-standing membranes, either by a fine etching strategy or by a potential shock approach.[18,19] and then are transferred to a desired substrate.[20-22] This leads to relatively high efficiencies when the tube layers (grown or transferred) on FTO are employed as a photo-anode in dye sensitized solar cells (DSSCs). *I.e.*, not only front- but also back-side illumination configuration of the semiconductor on the glass (*i.e.*, light passes through the substrate) can be used.[23,24] The advantages of a back-side arrangement of a photoelectrochemical cell are that the losses of light from a front-side illumination configuration due to the light absorption by a Pt counter electrode are avoided.[25] Furthermore, a light irradiation of the $TiO_2$ tube layers through the FTO glass is advantageous also because the majority of the charge carriers are generated close to the semiconductor/FTO interface and thus the charge collection efficiency by the FTO scaffold is substantially improved.[26,27]

Nevertheless, owing to their optical properties, FTO/glass substrates allow for only partial transmission of UV light, and especially wavelengths shorter than ~ 350 nm are substantially absorbed in FTO and hence do not reach the semiconductor.[28] Therefore, if one aims at studying in a back-side illumination configuration the photo-response of $TiO_2$ nanotubes under UV light irradiation, FTO should be replaced by a more suitable (*i.e.*, optically transparent) substrate, and quartz represent an ideal candidate for this purpose, showing almost 100% transmittance under irradiation with wavelengths as short as 250-350 nm.

In this work we illustrate an approach that allows for the growth of mechanically-stable $TiO_2$ nanotube layers directly on optically transparent quartz substrates. After crystallization into anatase $TiO_2$ and Pt co-catalyst deposition, these tube layers on quartz are studied as photocatalyst for $H_2$ production in both front-side irradiation and back-side illumination, and are characterized in view of their wavelength-dependent light transmission properties. We provide results on the effects of experimental parameters such as the irradiation configuration, the used light source and the amount of co-catalytic Pt, on the $H_2$ generation efficiency.

**Experimental**

Self-ordered vertically-aligned $TiO_2$ nanotube arrays were grown by anodization of Ti metal films (300 nm thick) that were deposited by electron beam evaporation on quartz slides (15 x 15 x 1 mm, GVB, Germany). The quartz substrates were previously cleaned by sonication in acetone, ethanol and deionized water (15 min for each step), and then dried in a $N_2$ stream. The evaporation of Ti onto the quartz slides was carried out with a deposition rate of 0.6 nm min$^{-1}$, at $5\times10^{-7} - 2\times10^{-6}$ mbar, using a PLS 500 Labor-system (Balzers-Pfeiffer).

The anodization of the Ti layers was performed in a two-electrode electrochemical O-ring cell (O-ring diameter of 10 mm) with a Pt plate as cathode. The electrolyte was composed of ethylene glycol with addition of 0.15 M $NH_4F$ and 3 vol% deionized water. The anodization of the Ti layers was conducted to completion, that is, until full conversion of the metal films into transparent $TiO_2$ nanotube layers, *i.e.*, no Ti metal was left underneath the nanotubes (for further experimental details see Results and Discussion). The anodization experiments were performed under potentiostatic conditions, that is, by applying a constant direct current (DC) potential of 60 V (no sweeping) provided by a Volcraft VLP 24 Pro DC power source. The resulting current density was recorded using a Keithley 2100 6½ Digit multimeter interfaced with a laptop.

The as-formed $TiO_2$ nanotubes on quartz were crystallized by annealing at 450°C, for 1 h, in air, using a rapid thermal annealer (Jipelec Jetfirst 100 RTA), with a heating and cooling rate of 30°C min$^{-1}$. Afterwards, Pt was deposited by plasma sputtering (EM SCD500, Leica, using a 99.99 % pure Pt target, Hauner Metallische Werkstoffe) onto the top of the crystalline tube layers in the form of films having various nominal thicknesses in the 0.5-10 nm range. The amount of Pt, *i.e.*, the nominal thickness of the Pt film, was *in situ* monitored with an automated quartz crystal film-thickness sensor. This approach allowed for a fine control of the Pt amount (*e.g.*, down to nominal thicknesses of 0.5 nm) and also ensured the reproducibility of the deposition. The sputtering was usually carried out at $10^{-2}$ mbar of Ar, by applying a current of 16 mA.

For the morphological characterization of the Pt-decorated $TiO_2$ nanotube layers, a field-emission scanning electron microscope (Hitachi FE-SEM S4800) was used. X-ray diffraction analysis (XRD, X'pert Philips MPD with a Panalytical X'celerator detector) using graphite monochromized CuKα radiation (wavelength 1.54056 Å) was performed for determining the crystalline structure of the samples. Their composition and the chemical state were characterized using X-ray photoelectron spectroscopy (XPS, PHI 5600, US) and peak positions were calibrated with respect to the Ti2p peak at 458 eV. Energy-dispersive X-ray spectroscopy (EDAX Genesis, fitted to SEM chamber) was also used for the chemical analysis of the anodic layers.

For the photocatalytic $H_2$ generation experiments, the samples were immersed into 20 vol% ethanol-water solutions within a sealed quartz tube. The ethanol-water solution (volume ~ 7 mL) and the cell head-space (volume ~ 8 mL) were purged with $N_2$ gas for 20 min prior to photocatalysis. Ethanol, as explained in more detail elsewhere,[429] was used as hole-scavenger, that is, it undergoes rapid oxidation (eventually towards $CO_2$) by reacting with photo-generated holes in $TiO_2$. This markedly suppresses the electron-hole recombination and makes conduction band electrons more readily available for the $H_2$ generation reaction.[30,31]

Two different light sources were used for the photocatalytic experiments: *i)* a CW-laser emitting UV light ($I_0$ = 7.8 mW cm$^{-2}$, λ = 266 nm, CryLas GmbH, Germany), and *ii)* an AM 1.5 solar simulator (300 W Xe, $I_0$ = 155 mW cm$^{-2}$, light spot size ~ 20 cm$^2$, Solarlight). The irradiation time was of 1 h for all the experiments. When using the UV light, the laser beam was expanded to a circle-shaped light spot of ~ 0.78 cm$^2$ to irradiate the full surface of the NT layers. In order to determine the amount of generated $H_2$, the gas that evolved under irradiation was accumulated within the headspace of the quartz reactor and was then analyzed by gas chromatography (using a GCMS-QO2010SE chromatograph, Shimadzu) withdrawing 200 μL samples with a gas tight syringe.

The light transmission properties of the TiO$_2$ NT films on quartz were also investigated. For this, we illuminated the NTs films with different monochromatic light sources in both front- and back-side irradiation configurations, and measured with a calibrated optical power meter (Newport, 1830-C) the intensity of transmitted light *I*, *i.e.*, the light that could pass through the nanotube films (the plain quartz substrates showed transparency at all the wavelengths). The used lasers had emission λ centered at 266, 325, 405, 473, and 633 nm, and emission intensity ($I_0$) of 7.8, 18.5, 18.2, 14.7, 0.086 mW cm$^{-2}$, respectively.

The Transmittance *T(%)* of the tube layers at the different irradiation λ was calculated as $T(\%) = \frac{I}{I_0} 100\%$. The results of the light transmittance measurements provide an overview of the light absorption properties of the nanotube films in dependence of the illumination wavelength.

**Results and Discussion**

In order to fabricate anodic TiO$_2$ NT layers that can be used in the standard front-side illumination configuration as well as in a back-side configuration (*i.e.*, illumination through the substrate), Ti metal layers were evaporated on optically transparent substrates and then were subjected to electrochemical anodization.

A key to achieve full anodization of the metal films and, at the same time, to avoid detachment (peeling-off) of the tube layers is a proper duration of the anodization experiments. For this, we monitored in real time the current *vs.* time profile and stopped the experiment at the suitable time.

In Fig. 1(a), a typical current density (*J*) *vs.* time profile is illustrated that presents three characteristic stages: *i)* after a current increase when the anodic bias is applied, *J* decreases in an exponential form due to the formation of the oxide barrier layer, by high-field oxidation,[15,32] *ii)* afterwards, *J* remains almost constant during the nanotube growth (under steady-state tube growth conditions[15,32]); *iii)* finally, *J* decreases to a very low current density values as the thin Ti layer is nearly fully consumed. In this phase, the continuous drop of *J* over time is ascribed to the fact that the front of oxide growth reaches the quartz first at the center of the anodized surface due to current distribution effects. In other words, the tubes grow slightly faster at the center of the anodic film as illustrated in the inset of Fig. 1a. As a consequence we observed that if the anodization is interrupted at "1" (see Fig. 1(a)), the growth of the nanotubes is complete only at the center of the anodized area, while at the edges there was still Ti metal left underneath the nanotubes. On the other hand, if the anodization is interrupted at "3", no Ti metal is left beneath the tubes (not even at the edges of the sample) but a large extent of oxide dissolution (etching-off) occurs at the center of the tube layer, leading to peeling-off (*i.e.*, or a bad adhesion) of the oxide film.

An optimal time for interrupting the anodization is thus at "2" (approximately when the current density reaches 50% of the steady-state phase) since: *i)* the amount of Ti metal left underneath the oxide at the edge of the sample is minimized, and it will be converted to oxide by the subsequent thermal treatment; and *ii)* the oxide-etching that takes place at the center of the sample occurs only to a small extent and does not affect the mechanical adhesion of the anodic film.

By this strategy, from the originally 300 nm-thick metallic Ti films, 700 nm-thick $TiO_2$ nanotube layers were formed (Fig. 1(b)). As shown in the SEM images in Fig. 1(c,d) and Fig. S1, these NTs are vertically aligned and show an average diameter of the mouth of ~ 30 nm (ascribed to the presence of the initiation layer[33]), and an average tube inner diameter of ~ 60 nm.

After fabrication, the as-formed amorphous TiO$_2$ NTs were converted into anatase phase by annealing in air at 450ºC. Then, Pt layers of various nominal thicknesses (in the 0.5-10 nm range) were sputter-deposited onto the anatase NTs.

From the optical pictures of the bare TiO$_2$ NTs shown in the insets of Fig. 1(a), one can see that the tube films are transparent. However, the films become darker in color and less transparent as thicker Pt films are deposited (see the inset in Fig. 1(d), and Fig S3 in the ESI). When sputtering relatively small noble metal amounts (e.g., ~ 1 nm-thick film) we found that small island-like Pt particles form at the top of the NTs (Fig. 1(c)), while for larger amount of sputtered metal (≥ 5 nm) Pt agglomerates atop the TiO$_2$ nanotubes (Fig. 1(d)), forming thick and dense layers (for further details see Table 1 and Fig. S2).

These structures, when characterized by XRD in view of their crystallographic properties, were found to be composed of anatase TiO$_2$ phase (with the characteristic anatase reflection peaking at 25.28°), and according to the annealing conditions weak rutile peak was detected (Fig. 2(a)). Also, the absence of a Ti metal peak confirms that the metal film is fully converted into oxide by the anodization step.

Both XRD patterns and EDX analyses do not show any Pt peak on the NTs when the sputtered Pt film is thinner than 1 nm (EDX data are compiled in Fig. S4(a)). However, for thicker films, the intensity of the Pt signals clearly increases with increasing its nominal thicknesses (characteristic XRD peaks of Pt are at ~ 40.0º and 46.5º, while its EDX signal peaks at ~ 2.1 eV), this confirming the presence of the noble metal and the accuracy and reproducibility of the Pt sputter-coating approach.

Additionally, XPS characterization was performed for all samples, particularly to verify the actual presence of Pt when thinner layers are sputter-deposited. The deposition of Pt (in the form of Pt$^0$) is confirmed for all the Pt-coated samples by the appearance of the Pt4f signal in the high resolution *spectra* (Fig. 2(b)), with the 4f$_{7/2}$ and 4f$_{5/2}$ signals peaking at ~ 71.8 and 75.2 eV, respectively. Moreover, from the XPS surveys (Fig. S4(b)) one can also see other characteristic Pt signal (*e.g.*, Pt3d peaking at ~ 314 eV), and the decrease of the Ti2p peak (~ 458 eV) with increasing the thickness of the sputtered Pt layer, this indicating a larger coverage of the $TiO_2$ surface with the noble metal.

The NT structures, decorated with different Pt amounts (*i.e.*, 0.5-10 nm) were then evaluated as photocatalytic films for the generation of $H_2$ from ethanol-water solutions. As anticipated above and illustrated in Fig. 3(a), the photocatalytic films were tested under both front- and back-side illumination, and by using different light sources (UV and simulated solar light).

From the photocatalytic data in Fig. 3, a first clear result is that regardless of the used light source, the plot of $H_2$ amount *vs.* the amount of co-catalyst shows specific trends for the two irradiation configurations.

Under front-side illumination, one can see that the amount of evolved $H_2$ increases steeply with increasing the Pt amount until 0.7 nm, and then it dramatically decreases for larger Pt amounts. In this configuration the largest $H_2$ production were measured with 0.7 nm Pt and are of ~ 23.5 and 8.8 $\mu L\ h^{-1}$, obtained under 266 nm and AM 1.5 simulated solar light irradiation, respectively.

On the contrary, under back-side illumination, the $H_2$ evolution efficiency increases monotonically with the Pt amount, although it substantially levels off when Pt > 1 nm. In this case, the largest $H_2$ amounts were measured with 10 nm of Pt and are ~ 8.5 and 4.5 $\mu L\ h^{-1}$, obtained under 266 nm and simulated solar light irradiation, respectively.

The reasons for the different trends are ascribed to the relative position of the co-catalyst with respect to the light irradiation pathway. In the front-side illumination, the majority of the charge carriers (*i.e.*, electron-hole pairs) are more likely generated at the top of the tubes and thus in the vicinity of the co-catalyst. Therefore, electrons have to migrate only over relatively short distances to reach the co-catalytic sites (the Pt nanoparticles) and generate $H_2$. Therefore, in this configuration the amount of Pt must be optimized for two factors: *i)* a maximized Pt decoration density, for a most efficient electron transfer towards the co-catalyst and *ii)* a minimized Pt amount regarding shading effects of the semiconductor (Pt absorbs and scatters the impinging light). Additionally, a large coverage of $TiO_2$ by Pt might hinder the transfer of valence band holes towards the environment, which results in the detrimental charge recombination (*i.e.*, a less efficient charge separation and transfer towards the solid/liquid interface).

The two latter *phenomena*, however, do not occur under back-side illumination and as a result the larger the Pt amount the higher is the $H_2$ evolution efficiency. In fact, since the light passes through the quartz substrate, the photon density impinging on the $TiO_2$ structures is not affected (shaded) by the amount of noble metal (excluding also back reflection effects). Also, a large coverage of the tube top with the co-catalyst is not expected to limit the hole transfer to the environment since the majority of the electron-hole pairs is generated at the tube bottom.

These factors on the relative position of the co-catalyst with respect to the light irradiation pathway provide also explanation for the larger $H_2$ evolution efficiency measured under front-side illumination compared to the back-side (~ 23.5 *vs.* 3.0 µL $h^{-1}$, respectively, under 266 nm light irradiation). In fact, as anticipated above the most photo-active part of the oxide layers under back-side illumination is the bottom of the tubes, and thus the electrons have to migrate over relatively long distances to reach the co-catalytic Pt at the tube top, this limits the $H_2$ generation. It is therefore clear that an optimized front-side illumination configuration represents a more effective experimental condition in view of a maximized photocatalytic $H_2$ generation.

Worth noting, these findings set apart from what is observed in photo-electrochemical (PEC) cells, that is that the illumination of the semiconductor through the transparent substrate (*i.e.*, back-side irradiation) is typically beneficial. Under open circuit condition, as studied here, a key factor is the electron diffusion through the semiconductor and towards the environment that is improved by the presence of the Pt decorations at its surface (as ascribed to the pinning of the semiconductor band levels at the interface with Pt[7,34,35]). Therefore, the overall efficiency of the photocatalytic $H_2$ evolution is strictly related to the effectiveness of the electron transfer towards the $TiO_2$/Pt/environment interface. In other words, a most efficient photocatalytic configuration should imply a direct irradiation of the $TiO_2$ NT/Pt/environment interface (optimized for minimized shading).

Fig. 3 shows that in our experiments, the 266 nm UV light leads to a higher $H_2$ generation efficiencies compared to solar light irradiation. This could trivially be ascribed to the fact that the UV integral spectral component of the simulated solar light (which is responsible for $TiO_2$ band gap excitation) is, in terms of power density, nearly the 5.4% of the full emission *spectrum*, that corresponds to a power of *ca.* 5.4 mW cm$^{-2}$, which is lower than the intensity of the monochromatic 266 nm laser (*e.g.*, ~ 7.8 mW cm$^{-2}$).

However, it is interesting to evaluate the photocatalytic results in terms of evolved $H_2$ amount normalized *vs.* the UV irradiation power (under front-side illumination and with the optimized Pt amount, *i.e.*, 0.7 nm): the monochromatic 266 nm light (1.4 times higher photon flux than the UV component of the solar simulator) leads to a $H_2$ production that is 2.7 time higher than that obtained under AM 1.5 illumination. This observation suggests that the difference in $H_2$ generation is also related to the smaller penetration depth in the tubes of the shorter UV wavelengths (266 nm), so that the electron diffusion has to occur over shorter distances to reach the co-catalytic Pt at the top of the tubes. Conversely, the polychromatic UV component of the AM 1.5 simulated solar light (composed of UV wavelength that are closer to the $TiO_2$ absorption threshold, *i.e.*, ~ 400 nm) has larger penetration depth, and hence the generated carriers may experience a large extent of recombination since they have to migrate over longer pathways to reach the $TiO_2$/Pt/environment interface.[36,37] This interpretation is supported by the fact that in the back-side irradiation configuration the $H_2$ generation rates under 266 nm and under solar light irradiation are similar. In fact under back-side irradiation the lower UV power density of the simulated solar light is compensated by the larger penetration depth of its photons, and owing to the irradiation configuration, the charge carriers are thus generated closer to the $TiO_2$/Pt/environment interface and their migration towards the Pt co-catalyst is facilitated.

The bare and Pt-decorated (0.7 nm) anatase $TiO_2$ nanotube layers on quartz were also investigated in view of their wavelength-dependent light transmission features, to provide complementary information to the photocatalytic results. From the data compiled in Fig. 4 one can notice that regardless of the presence of Pt, the general trend of light transmission over the irradiation wavelength does not depend on the light illumination pathway (front- *vs.* back-side irradiation). The low transmittance observed under UV light irradiation ($\lambda < 405$ nm) and its significant increase at longer wavelength (*T(%)* reaches values of *ca.* 90% under 633 nm light irradiation) can directly be ascribed to the bandgap of anatase $TiO_2$ (*i.e.*, super- *vs.* sub-bandgap illumination).

These transmission data, being measured under monochromatic irradiation at different wavelength, may also be compared to literature data on the wavelength-dependent light absorption coefficients ($\varepsilon_{TiO_2,\lambda}$) of anatase $TiO_2$ nanotube layers.[38]

While in the literature for NT membranes values of $\varepsilon_{TiO_2}$ of 0.39 and 0.43 µm$^{-1}$ are reported under super-bandgap irradiation at 266 and 365 nm, respectively, an evaluation of our transmission data yields $\varepsilon_{TiO_2}$ values of ~ 2.8 µm$^{-1}$. This higher light absorption coefficient of the layers studied here may be the consequence of the fact that tube layers fabricated from Ti films e-beam evaporated on quartz not only carry the "initiation layer" on top of the tubes (shown in Fig. 1 and Fig S1 in the ESI) but also show the typical bottom-close morphology – *i.e.*, likely, these structural features, that are absent in the case of the tube membranes studied in literature,[38] contribute to a larger light absorption of the layers studied here.

Furthermore, a direct comparison between the data in Fig. 4(a,b) shows the effects of the presence of Pt on the light absorption properties of the photocatalysts. As expected, upon deposition of a 0.7 nm-thick Pt film, the light transmission of the oxide films is noticeably reduced at all the irradiation wavelengths. These results not only are in line with the darkening of the NT layers observed upon Pt deposition (shown in the optical images in Fig. 1 and Fig. S3), but also confirm that the presence of Pt in small amounts at the tube top already shades considerably the underneath semiconductor.

**Conclusions**

This work shows that for photocatalytic films based on Pt-decorated $TiO_2$ nanotube layers on transparent substrates, the irradiation configuration and the amount of co-catalyst are relevant parameters to be optimized in view of maximizing the $H_2$ evolution. Regarding the Pt loading not only the amount and distribution density affect the photocatalytic efficiency (via beneficial co-catalytic *vs.* shading effects) but also the light illumination pathway. A most efficient photocatalytic configuration was found using front-side irradiation with short UV wavelength (266 nm) and a nominal loading of 0.7 nm Pt on top of the tubes.

**Acknowledgements**

The authors would like to acknowledge the ERC, the DFG, and the DFG "Engineering of Advanced Materials" cluster of excellence and DFG "funcos" for financial support. This project was funded by the Deanship of Scientific Research (DSR), King Abdulaziz University, under grant no. 16-130-36-HiCi. H. Hildebrand is acknowledged for valuable technical help. O. Pfoch and M. Licklederer are acknowledged for the deposition of Ti films on the quartz slides.

# References

1. Fujishima, A. & Honda, K. Electrochemical photolysis of water at a semiconductor electrode. *Nature* **238,** 37–38 (1972).

2. Kawai, T. & Sakata, T. Conversion of carbohydrate into hydrogen fuel by a photocatalytic process. *Nature* **286,** 474–476 (1980).

3. Hoffmann, M. R., Martin, S. T., Choi, W. & Bahnemannt, D. W. Environmental Applications of Semiconductor Photocatalysis. *Chem. Rev.* **95,** 69–96 (1995).

4. Fox, M. A. & Dulay, M. T. Heterogeneous Photocatalysis. *Chem. Rev.* **93,** 341–357 (1993).

5. Palmisano, G., Augugliaro, V., Pagliaro, M. & Palmisano, L. Photocatalysis: a promising route for 21st century organic chemistry. *Chem. Commun. (Camb).* **33,** 3425–37 (2007).

6. Tripathy, J., Loget, G., Altomare, M. & Schmuki, P. Photocatalytic Oxidation of Benzyl Alcohol to Benzaldehyde Under Visible Light. **16,** 1–6 (2016).

7. Linsebigler, A. L. *et al.* Photocatalysis on TiO2 Surfaces: Principles, Mechanisms, and Selected Results. *Chem. Rev.* **95,** 735–758 (1995).

8. Ghicov, A., Schmidt, B., Kunze, J. & Schmuki, P. Photoresponse in the visible range from Cr doped TiO2 nanotubes. *Chem. Phys. Lett.* **433,** 323–326 (2007).

9. Das, C., Roy, P., Yang, M., Jha, H. & Schmuki, P. Nb doped TiO(2) nanotubes for enhanced photoelectrochemical water-splitting. *Nanoscale* **3,** 3094–3096 (2011).

10. Paramasivam, I., Nah, Y. C., Das, C., Shrestha, N. K. & Schmuki, P. WO3/TiO2 nanotubes with strongly enhanced photocatalytic activity. *Chem. - A Eur. J.* **16,** 8993–8997 (2010).

11. Salazar, R. *et al.* Use of Anodic TiO\n 2\n Nanotube Layers as Mesoporous Scaffolds for Fabricating CH\n 3\n NH\n 3\n PbI\n 3\n Perovskite-Based Solid-State Solar Cells. *ChemElectroChem* n/a–n/a (2015). doi:10.1002/celc.201500031

12. Liu, N. *et al.* Black TiO2 Nanotubes: Cocatalyst-Free Open-Circuit Hydrogen Generation. *Nano Lett.* **14,** 3309–3313 (2014).

13. Mohammadpour, F. *et al.* Enhanced performance of dye-sensitized solar cells based on TiO\n 2\n nanotube membranes using an optimized annealing profile. *Chem. Commun.* **51,** 1631–1634 (2015).

14. Lee, K., Mazare, A. & Schmuki, P. One-dimensional titanium dioxide nanomaterials: Nanotubes. *Chem. Rev.* **114,** 9385–9454 (2014).

15. Roy, P., Berger, S. & Schmuki, P. TiO2 Nanotubes: Synthesis and Applications. *Angew. Chemie Int. Ed.* **50,** 2904–2939 (2011).
14

16. Leenheer, A. J., Miedaner, A., Curtis, C. J., van Hest, M. F. a. M. & Ginley, D. S. Fabrication of nanoporous titania on glass and transparent conducting oxide substrates by anodization of titanium films. *J. Mater. Res.* **22,** 681–687 (2007).

17. Stergiopoulos, T. *et al.* Dye-sensitization of self-assembled titania nanotubes prepared by galvanostatic anodization of Ti sputtered on conductive glass. *Nanotechnology* **20,** 365601 (2009).

18. Jo, Y., Jung, I., Lee, I., Choi, J. & Tak, Y. Fabrication of through-hole TiO2 nanotubes by potential shock. *Electrochem. commun.* **12,** 616–619 (2010).

19. Chen, Q. & Xu, D. Large-Scale, Noncurling, and Free-Standing Crystallized $TiO_2$ Nanotube Arrays for Dye-Sensitized Solar Cells. *J. Phys. Chem. C* **113,** 6310–6314 (2009).

20. Cha, G., Lee, K., Yoo, J., Killian, M. S. & Schmuki, P. Topographical study of TiO2 nanostructure surface for photocatalytic hydrogen production. *Electrochim. Acta* **179,** 423–430 (2015).

21. Mohammadpour, F. *et al.* Comparison of Anodic $TiO_2$-Nanotube Membranes used for Frontside-Illuminated Dye-Sensitized Solar Cells. *ChemElectroChem* **2,** 204–207 (2015).

22. Mohammadpour, F. *et al.* High-temperature annealing of $TiO_2$ nanotube membranes for efficient dye-sensitized solar cells. *Semicond. Sci. Technol.* **31,** 014010 (2016).

23. Mor, G. K., Varghese, O. K., Paulose, M. & Grimes, C. a. Transparent highly ordered TiO2 nanotube arrays via anodization of titanium thin films. *Adv. Funct. Mater.* **15,** 1291–1296 (2005).

24. Mohammadpour, F. *et al.* Enhanced performance of dye-sensitized solar cells based on TiO2 nanotube membranes using an optimized annealing profile. *Chem. Commun.* **51,** 1631–1634 (2015).

25. Li, L.-L., Chen, Y.-J., Wu, H.-P., Wang, N. S. & Diau, E. W.-G. Detachment and transfer of ordered TiO2 nanotube arrays for front-illuminated dye-sensitized solar cells. *Energy Environ. Sci.* **4,** 3420 (2011).

26. Lee, K., Kirchgeorg, R. & Schmuki, P. Role of Transparent Electrodes for High Efficiency $TiO_2$ Nanotube Based Dye-Sensitized Solar Cells. **118,** 16562–16566 (2014).

27. So, S., Kriesch, A., Peschel, U. & Schmuki, P. Conical-shaped titania nanotubes for optimized light management in DSSCs reach back-side illumination efficiencies > 8%. *J. Mater. Chem. A* **3,** 12603–12608 (2015).

28. Li, Q. *et al.* ZnO nanoneedle/H2O solid-liquid heterojunction-based self-powered ultraviolet detector. *Nanoscale Res. Lett.* **8,** 415 (2013).

29. Bamwenda, G. R., Tsubota, S., Nakamura, T. & Haruta, M. Photoassisted hydrogen production from a water-ethanol solution: a comparison of activities of AuTiO2 and PtTiO2. *J. Photochem. Photobiol. A Chem.* **89,** 177–189 (1995).

30. Fujishima, A., Zhang, X. & Tryk, D. a. TiO2 photocatalysis and related surface phenomena. *Surf. Sci. Rep.* **63,** 515–582 (2008).

31. Murdoch, M. *et al.* The effect of gold loading and particle size on photocatalytic hydrogen production from ethanol over Au/TiO$_2$ nanoparticles. *Nat. Chem.* **3,** 489–492 (2011).

32. Macak, J. M. *et al.* TiO2 nanotubes: Self-organized electrochemical formation, properties and applications. *Curr. Opin. Solid State Mater. Sci.* **11,** 3–18 (2007).

33. Berger, S., Macak, J. M., Kunze, J. & Schmuki, P. High-Efficiency Conversion of Sputtered Ti Thin Films into TiO2 Nanotubular Layers. *Electrochem. Solid-State Lett.* **11,** C37 (2008).

34. Basahel, S. N. *et al.* Self-decoration of Pt metal particles on TiO2 nanotubes used for highly efficient photocatalytic H2 production. *Chem. Commun.* **50,** 6123–5 (2014).

35. Kamat, P. V. Photophysical, Photochemical and Photocatalytic Aspects of Metal Nanoparticles. *J. Phys. Chem. B* **106,** 7729–7744 (2002).

36. Lynch, R. P., Ghicov, A. & Schmuki, P. A Photo-Electrochemical Investigation of Self-Organized TiO[sub 2] Nanotubes. *J. Electrochem. Soc.* **157,** G76 (2010).

37. Jennings, J. R., Ghicov, a, Peter, L. M., Schmuki, P. & Walker, a B. Dye-Sensitized Solar Cells Based on Oriented TiO2 Nanotube Arrays: Transport, Trapping, and Transfer of Electrons - Journal of the American Chemical Society (ACS Publications). *J. Am. Chem. Soc.* 13364–13372 (2008). doi:10.1021/ja804852z

38. Cha, G., Schmuki, P. & Altomare, M. Free standing membranes to study the optical properties of anodic TiO2 nanotube layers. *Chem. - An Asian J.* n/a–n/a (2016). doi:10.1002/asia.201501336

**Table**

| Nominal thickness of the sputtered Pt layer | Size of the Pt nanoparticles |
|---|---|
| Pt-0 | - |
| Pt - 0.7 nm | 3 ~ 8 nm |
| Pt - 1 nm | 4 ~ 16 nm |
| Pt - 5 nm | Thin Pt layer |
| Pt - 10 nm | Thick Pt layer |

*Table 1* Size of Pt on TiO$_2$ NTs as a function of Pt layer thickness.

**Figures**

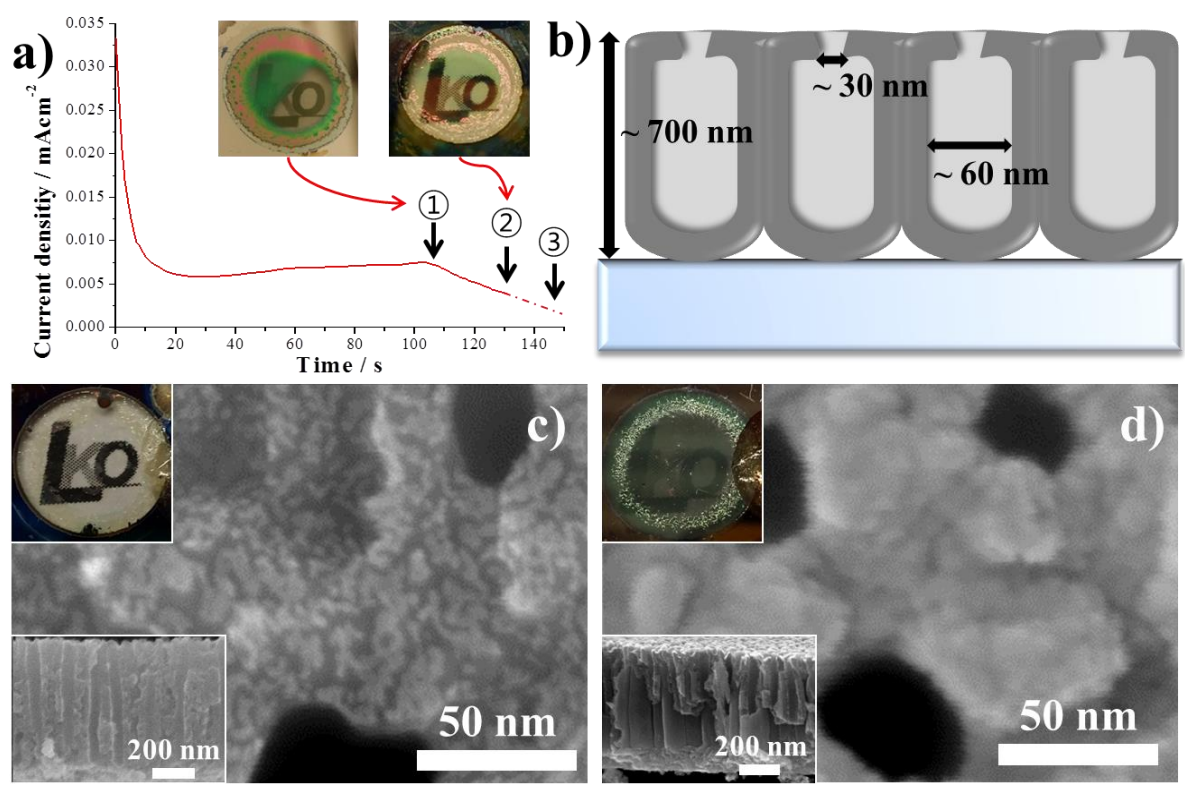

*Figure 1* a) I-T curve during anodizing of Ti sputtered on quartz, at 60V - optical pictures after anodization (inset), b) schematic representation of anodic $TiO_2$ NTs on quartz substrate, SEM and optical images of Pt/$TiO_2$ NTs with a Pt layer thickness of: c) 1 nm; d) 10nm. 1 nm thick Pt shows the clearest LKO-logo behind the $TiO_2$ NTs sample; in contrast, 10nm thick Pt decorated sample shows low transparency with dark green color.
(Morphological characterization of 0.5 nm thick Pt was similar with that of 0.7 thick Pt, not shown here.)

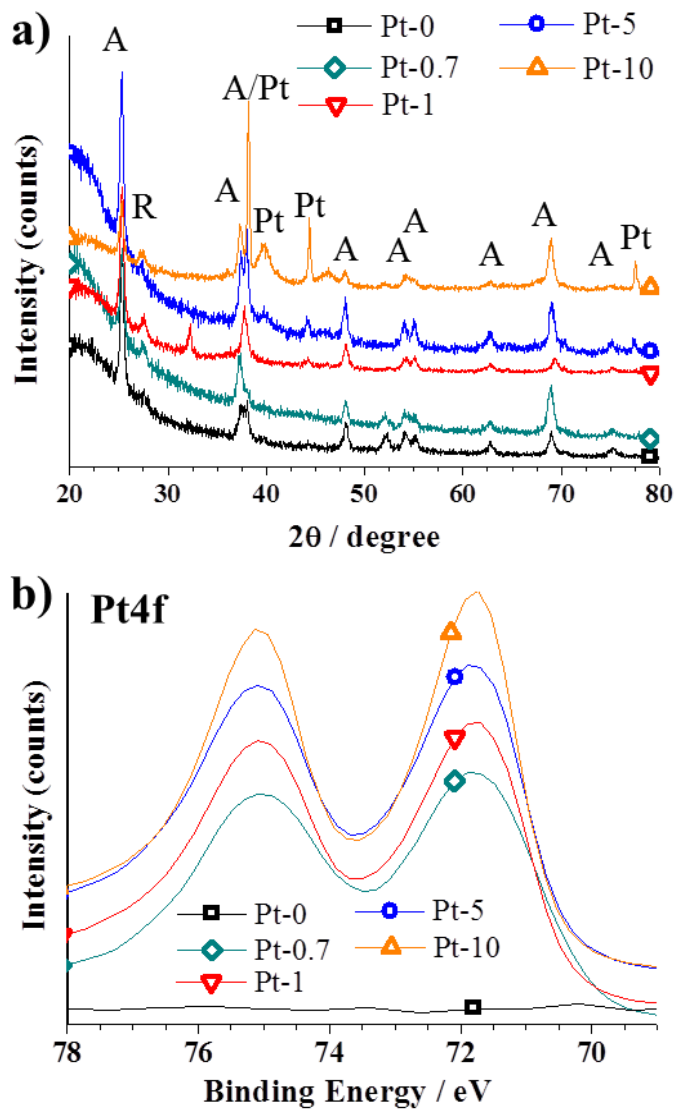

*Figure 2* a) XRD patterns, b) Pt 4f XPS spectrum for bare annealed TiO$_2$ nanotubes, and Pt/TiO$_2$ NTs with 0.7 nm, 1 nm, 5 nm, 10 nm thick Pt layers. A = Anatase, Si = Silicon, Q = Quartz, Pt = Platinum, Ti = Titanium, O = Oxygen, N = Nitrogen. (Chemical characterization of 0.5 nm thick Pt was similar with that of 0.7 thick Pt, not shown here.)

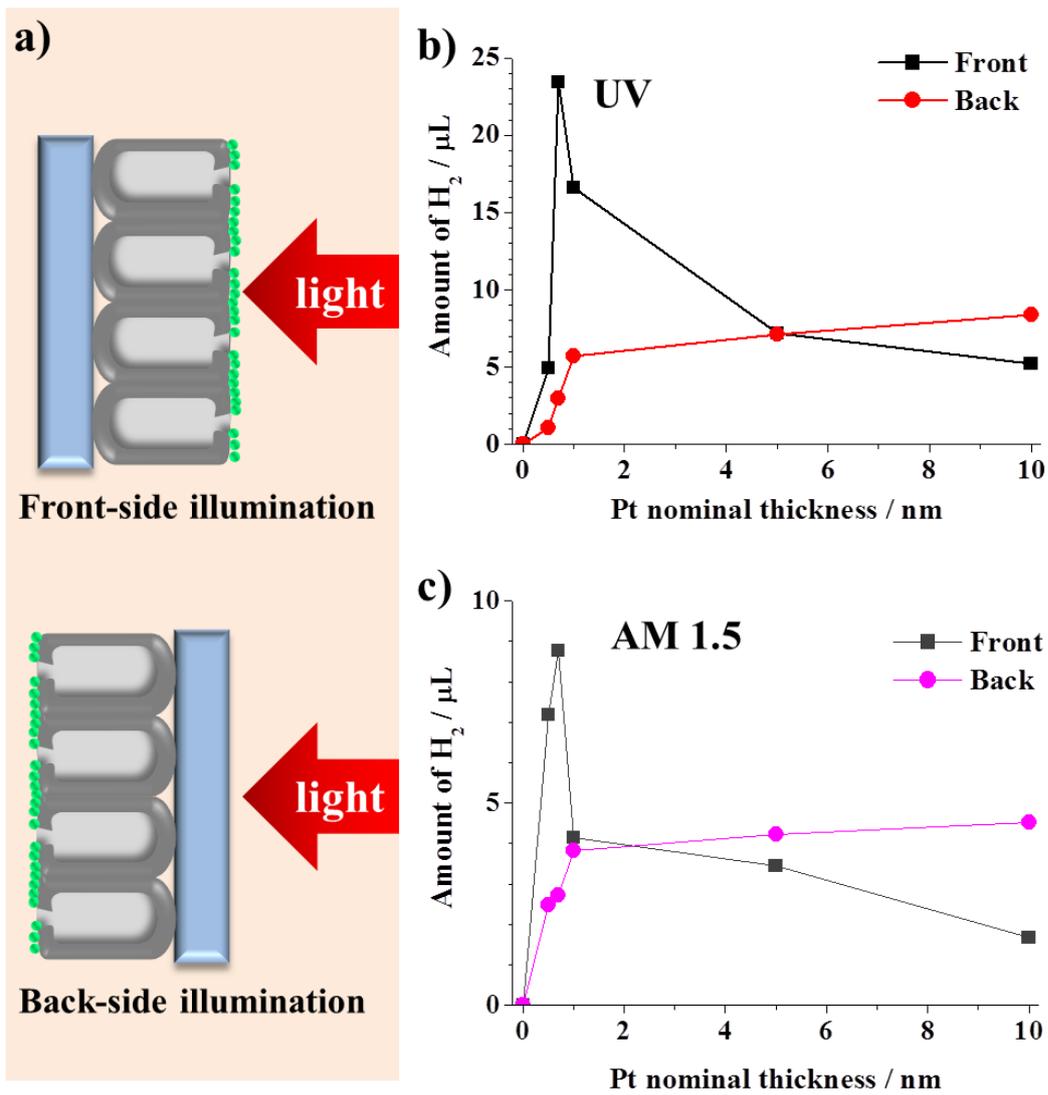

*Figure 3* a) schematic representation of the induced light's path: upper part, front-side illumination; lower part, back-side illumination. Photocatalytic $H_2$ production for Pt/$TiO_2$ NTs, as a function of the thickness of the Pt layer, for 1 h illumination under: b) UV light and c) AM 1.5 light.

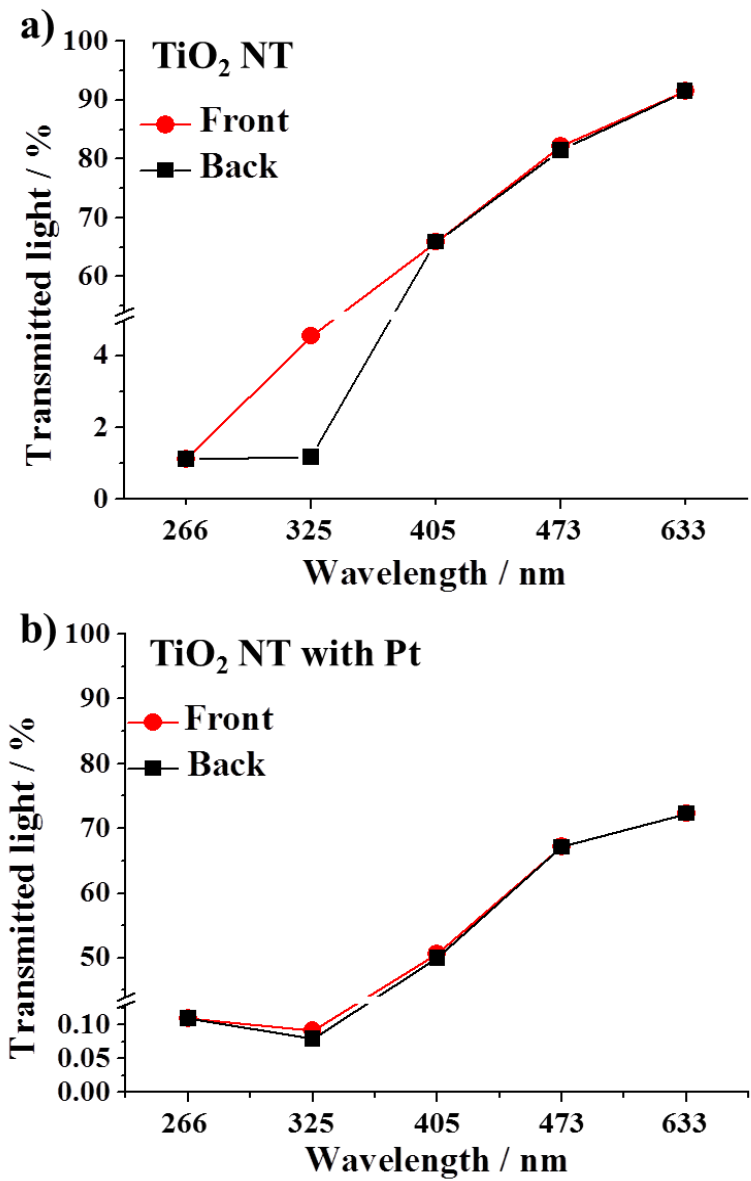

*Figure 4* The transmitted light intensity, in front- and back-side illumination, after passing through: a) annealed TiO$_2$ NTs on quartz, b) 0.7 nm thick Pt decorated annealed TiO$_2$ NTs on quartz.